\documentclass[useAMS,usenatbib]{mn2e}
\usepackage[english]{babel}
\usepackage{graphicx}
\usepackage{amsmath}
\usepackage{amssymb}
\usepackage{epsf}
\usepackage{bm}
\usepackage[dvips]{color}

\begin{document}

\newcommand{\rate}{\mbox{$\mathrm{erg~cm^{-3}~s^{-1}}$}}
\newcommand{\wrate}{\mbox{$\mathrm{erg~s^{-1}}$}}
\newcommand{\gcc}{\mbox{$\mathrm{g~cm^{-3}}$}}
\newcommand{\beq}{\begin{equation}}
\newcommand{\eeq}{\end{equation}}
%
%
\title[Thermal emission of neutron stars with internal heaters]
{Thermal emission of neutron stars with internal heaters}

\author[A.~D.~Kaminker et al.]
{ A.~D.~Kaminker$^{1}$,
          A.~A.~Kaurov$^{2}$,
          A.~Y.~Potekhin$^{1,3,4}$ and
          D.~G.~Yakovlev$^{1}$\\
$^{1}$Ioffe Physical-Technical Institute,
Politekhnicheskaya 26, Saint Petersburg, 194021, Russia\\
$^{2}$Department of Astronomy and Astrophysics, University of
Chicago, Chicago IL 60637, USA\\
$^{3}$Central Astronomical Observatory of RAS at Pulkovo,
Pulkovskoe Shosse 65, Saint Petersburg, 196140, Russia\\
$^{4}$Isaac Newton Institute of Chile, St.~Petersburg
Branch, Russia}
%
\date{Accepted 2013 xxxx. Received 2013 xxxx;
in original form 2013 xxxx}

\pagerange{\pageref{firstpage}--\pageref{lastpage}} \pubyear{2012}

\maketitle

\label{firstpage}

\begin{abstract}
Using 1D and 2D cooling codes we study thermal emission from neutron
stars with steady state internal heaters of various
intensities and geometries
(blobs or spherical layers) located at different depths in
the crust. The generated heat tends to propagate radially, from the
heater down to the stellar core and up to the surface; it is also
emitted by neutrinos. In local regions near the heater the results
are well described with the 1D code. The heater's region projects
onto the stellar surface forming a hot spot.  There are two heat
propagation regimes. In the first, \emph{conduction outflow} regime
(realized at heat rates $H_0 \lesssim
10^{20}$~erg~cm$^{-3}$~s$^{-1}$ or temperatures $T_\mathrm{h} \lesssim
10^9$~K in the heater) the thermal surface emission of the star
depends on the heater's power and neutrino emission in the
stellar
core. In the second, \emph{neutrino outflow} regime
($H_0 \ga 10^{20}\, \rate$ or $T_\mathrm{h} \ga 10^9$~K)
the surface thermal emission
becomes independent of heater's
power and the physics of the core. The largest (a few per cent)
fraction of heat power is carried to the surface if the heater is in
the outer crust and the heat regime is intermediate.  The results
can be used for modeling
young cooling neutron stars (prior to the
end of internal thermal relaxation), neutron stars in X-ray
transients, magnetars and high-$B$ pulsars,
as well as merging neutron stars.
\end{abstract}

\begin{keywords}
dense matter --- stars: neutron -- neutrinos.
\end{keywords}

\section{Introduction}
\label{s:introduc}

Neutron stars manifest themselves in
different ways (e.g. \citealt*{hpy07}). They are born hot in
supernova explosions but gradually cool down due to neutrino
emission from the entire stellar body and due to photon
emission from the surface. Thermal radiation from isolated cooling neutron
stars carries important information on internal structure of these
stars. Moreover, different mechanisms of extra energy release
can operate inside
the neutron stars  (e.g.,
\citealt*{pgw06}). For instance, an extra heating can be provided by
viscous friction in the presence of differential rotation (e.g.,
\citealt*{csy13}), slow chemical equilibration of the star in the
course of its evolution \citep{pr10}, Ohmic decay of internal
magnetic fields (in ordinary neutron stars and magnetars; e.g.,
\citealt*{viganoetal13} and references therein), nuclear reactions in
deep layers of the star's crust \citep*{hz90,hz08,bbr98},
or glitches (e.g., \citealt{espinozaetal11}).

In this paper we study possible manifestations of the
internal heaters in producing thermal surface radiation. We
will model the heaters phenomenologically as some hot
quasistationary heat sources of different size and
intensity, located in various places of the neutron star crust,
and see how much heat can diffuse to the surface and be
emitted as thermal radiation. The results will be helpful
for constraining the properties of the heater and its dense
environment from observations of thermal radiation from
neutron stars.

We have already studied the formulated problem in papers
devoted to magnetars, i.e., neutron stars with very strong magnetic fields $B \gtrsim
10^{14}$~G
\citep{kypssg06,kpyc09} (hereafter Papers I and II, respectively).
The aim was to explain quasistationary
thermal emission of magnetars
(e.g., \citealt{m08,m13,ok13}) by the presence
of internal heaters. To this aim, we have used our
generally relativistic 1D cooling code \citep*{gyp01} with
a phenomenological
spherically symmetric heat layer in the
neutron star crust. The results were summarized
in Paper~I for
heat-blanketing envelopes made of iron
and in Paper~II for magnetar models with
accreted heat-blanketing
envelopes. The description of heat transport in the stellar
interior (under the heat blanketing envelope, at densities $\rho
\gtrsim \rho_\mathrm{b} \sim 10^{10}$ g~cm$^{-3}$) was
approximate,
because the temperature distribution in the interior was treated as
spherically symmetric and the anisotropy of heat transport was
neglected.

This paper is different from the previous ones in two respects. First,
we supplement our 1D calculations by the calculations made with our
new 2D code. This allows us to consider axially symmetric heaters
and temperature distributions in the stellar interior. Second, we
mainly focus on neutron stars without magnetic fields
or with fields
$B \lesssim 10^{12}$~G,
which weakly affect thermal structure
and evolution  of the stars
(e.g., \citealt{yp04}).
In this case our
scheme of solving the heat transport problem in the stellar
interior is more robust than in Papers I and~II. Preliminary
results of the present study have been published by
\citet{kkpy12}. We show that many observational manifestations of
internal steady state heaters obtained for magnetars are
also valid for
ordinary neutron stars.

\section{Physics input}
\label{s:physics}

We calculate thermal radiation from neutron stars with internal
heaters using two cooling codes. First, we employ our usual 1D fully
relativistic cooling code (\citealt*{gyp01}; also see Papers~I and
II) with a spherically symmetric heater. Second, we use a new
simplified 2D cooling code \citep{kkpy12} with an axially symmetric
heater,
like a hot blob within the neutron star crust.
The two
codes allow us to follow the cooling more reliably.

Both codes simulate cooling of an initially hot star via neutrino
emission from the entire stellar body and via thermal emission of
photons from the surface. To facilitate calculations we use the
standard procedure of dividing the star in the bulk interior and a
thin outer heat-blanketing envelope \citep*{gpe83}. The envelope
extends from the radiative surface to the layer of the density
$\rho_\mathrm{b} = 10^{10}$~\gcc; typically it has a thickness $\sim
100$ m and mass $\lesssim10^{-6}\,M_\odot$. We consider the
ground-state composition of the matter according to \citet*{rhs06}.
The cooling results for this composition are almost the same as
those obtained with the previous models of the ground state matter
(e.g., \citealt{hpy07}) or purely iron heat blankets.  The
neutron drip density in the crust for all equations of state (EOSs)
used in this paper is $\rho_d \approx 4 \times 10^{11}$
g~cm$^{-3}$.

The internal structure of neutron stars can be regarded as
temperature-independent (e.g., \citealt{hpy07}). The 1D code
solves the \emph{thermal balance} and \emph{thermal transport}
equations in the entire bulk interior
(the crust and the core) in a
spherically symmetric approximation,
as described in \citet{gyp01}.

Our new 2D code solves
the thermal balance and thermal transport equations
in the
\emph{axially symmetric} approximation
 (see Sect.\,3.1 of \citealt*{apm08})
in the bulk of the crust
in a locally flat reference frame.
In the latter case we
calculate all relevant quantities
(temperature $T$, heat flux $\bm{F}$,
neutrino emissivity $Q_\nu$ [\rate], etc.)
as functions of radial coordinate $r$, Schwarzschild time $t$, and
a polar angle $\theta$
with respect to the
symmetry axis. The effects of General
Relativity are taken into account by
redshifting the results for a
distant observer.

The stellar core in the 2D code is treated as
isothermal and
included approximately by introducing the crust-core boundary
conditions. These conditions require the temperature over the boundary to be isothermal,
$T=T_{\rm cc}$, and the generally relativistic equation
of the core cooling to be satisfied, $d{T}_{\rm cc}^\infty/dt=
-L_{\nu \rm core}^\infty({T}_{\rm cc}^\infty)
/C^\infty_\mathrm{core}({T}_{\rm cc}^\infty)$.
Here $L^\infty_{\nu \rm core}({T}_{\rm cc}^\infty)$ and
$C^\infty_\mathrm{core}({T}_{\rm cc}^\infty)$
are, respectively, the
neutrino luminosity and heat capacity of the
isothermal core redshifted for a distant observer,
e.g., \citet{gyp01};  ${T}_{\rm cc}^\infty$ is the redshifted ${T}_{\rm cc}$.
The function $L_{\nu \rm core}^\infty({T}_{\rm cc}^\infty)
/C_{\rm core}^\infty({T}_{\rm cc}^\infty)$ is calculated for a given neutron star model in a standard way. Similar simplified cooling models have been introduced earlier
(e.g., \citealt{Fortin-ea10}).

The specific heat capacity $c_v$ in the core is calculated as the sum
of contributions of strongly degenerate relativistic
electrons and nucleons according to \citet*{yls99}.
In the crust, the contribution of the lattice of atomic
nuclei \citep*{BaikoPY01} is added to those of free neutrons
and electrons. The heat conductivity in the crust is
mainly regulated by electron-ion scattering. For the
isotropic case considered here, it is
given in \citet{pbhy99}.
The neutrino emissivity $Q_\nu$ is taken from
\citet{yakovlevetal01}. Unlike the earlier works (e.g.,
\citealt{gpe83,pcy97,pcyg03}),
we include
neutrino
emission in the heat-blanketing envelope,
which can be important at high effective temperatures.
In this case,
the radial flux $F_r$ is not
constant through the heat-blanketing envelope. Therefore,
both the effective surface temperature $T_\mathrm{s}$ and
the flux value $F_r=F_\mathrm{b}$ at $\rho_\mathrm{b}$,
have been simultaneously determined
for a given temperature $T_\mathrm{b}$
by integration of a system of
stationary equations of hydrostatic balance, thermal
balance and thermal transport in the blanketing
envelope, as described in \citet{pcy07}.
The boundary conditions at $\rho=\rho_\mathrm{b}$ implied the continuity
of $T$ and $F_r$.

\begin{table}
\caption[]{Masses $M$, radii $R$ and central densities in
units of $10^{14}$~\gcc, $\rho_\mathrm{c14}$, for neutron star
models with the toy-model EOS SC+HHJ (for short, HHJ) and realistic
EOS BSk21 (for short, BSk); see text. }
\label{tab:models}
\begin{center}
\begin{tabular}{ l c@{\hspace{2ex}}c c@{\hspace{2ex}}c c@{\hspace{2ex}}c}
\hline 
Star model & \multicolumn{2}{c}{$M/M_\odot$} &
\multicolumn{2}{c}{$R$ (km)} &
\multicolumn{2}{c}{$\rho_\mathrm{c14}$}
   \\
\hline \hline
& HHJ & BSk & HHJ & BSk & HHJ & BSk \\
Maximum mass & 2.16 & 2.28  &  10.84 & 11.07 & 24.5 & 22.9
\\
Fast cooler & \multicolumn{2}{c}{1.85}    &  12.32 & 12.46 & 11.34 & 9.98
  \\
 Durca onset & 1.77 & 1.57   &  12.46 & 12.58 & 10.50 & 8.09
  \\
Standard cooler     & \multicolumn{2}{c}{1.4}      &  12.74 & 12.57 &
7.78 & 7.30
 \\
\hline 
\end{tabular}
\end{center}
\end{table}

As will be shown below, our basic results depend on the
employed EOS only weakly. For this reason, most of our calculations
are performed with the use of a toy-model EOS, following Papers I
and II. Specifically, in the neutron-star core we employ the simple
parametrization of the energy  $\mathcal{E}$ per nucleon that was
constructed by \citet{hhj99} (hereafter HHJ),
\beq
   \mathcal{E} = \mathcal{E}_0 u\,\frac{u-2-s}{1+su}
           + S_0 u^\gamma (1-2x_\mathrm{p})^2.
\label{HHJ}
\eeq
Here, $u=n/n_0$; $n$ is the baryon density;
$n_0=0.16$ fm$^{-3}$, $\mathcal{E}_0$ and $S_0$ are, respectively,
the baryon density, nucleon energy, and symmetry energy
at saturation; $x_\mathrm{p}$ is the proton fraction (among baryons);
$s$ and $\gamma$ are additional parameters. Following HHJ, we keep
$\mathcal{E}_0=15.8$~MeV and $S_0=32$ MeV, but treat $s$ and
$\gamma$ as free parameters. We denote this parametrization
as HHJ($s,\gamma$). In the attempt to fit the EOS by
\citet*{apr98}, HHJ took $s=0.2$ and $\gamma=0.6$.
\citet{gusakovetal05} studied cooling of neutron stars with
the EOSs HHJ(0.2,$\gamma$) for three values of $\gamma=0.6$,
0.575 and 0.643, and denoted these models as APR I, II and
III, respectively. The latter model was also adopted in Papers I
and II.

\begin{figure}
\includegraphics[width=\columnwidth]{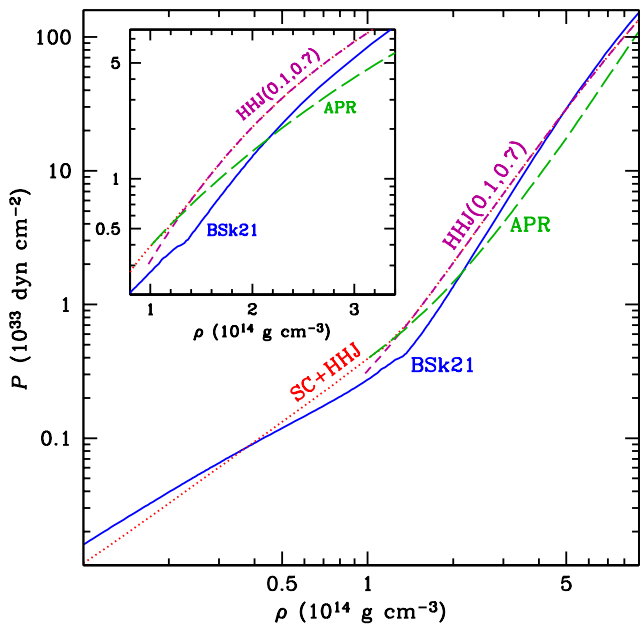}
\caption{EOS models for the inner crust and the core: APR
(long-dashed line) and HHJ(0.1,0.7) (short-dashed line) in the
core, interpolated SC+HHJ (dotted line) and unified BSk21
(solid line) in the core and crust. The inset shows
the same EOSs in a restricted range of densities.}
\label{fig:EOS}
\end{figure}

However, the HHJ
parametrization
is inaccurate in
reproducing the original EOS of
\citet{apr98}. For instance, all three HHJ parametrizations
of \citet{gusakovetal05} give maximum neutron-star masses
$M_\mathrm{max}\approx1.9\,M_\odot$,
significantly lower than the value
$2.2\,M_\odot$ for the true APR EOS.
It is lower than the
value $M=2.0\,M_\odot$
obtained in the
modern observations
of neutron stars
\citep{demorestetal10,antoniadisetal13}.
Therefore,
in the present
study we use the model HHJ(0.1,0.7) with
$M_\mathrm{max}=2.16\,M_\odot$ (Table \ref{tab:models}). In
the inner crust at
$\rho_\mathrm{b} < \rho < \rho_\mathrm{cc}$,
where $\rho_\mathrm{cc}$ refers to the
crust-core boundary
(e.g.,
$\rho_\mathrm{cc}=1.3\times10^{14}$~\gcc{}
according to \citealt{Pearson-ea12}), we
match
HHJ(0.1,0.7)
with the ``smooth composition'' (SC) EOS
described by \citet{hpy07}.
In practical calculations, we
apply a spline-like matching between HHJ(0.1,0.7) and SC EOS
in a range of densities near $\rho_\mathrm{cc}$.
We denote this combined EOS as ``SC+HHJ''
(see Fig.~\ref{fig:EOS}).

In order to check the sensitivity to the EOS model, we
compare the results obtained with the SC+HHJ EOS
and analogous results obtained with the analytical
representation of the realistic BSk21 EOS. This  model belongs  to
a family of EOSs based on nuclear energy-density
functionals, labelled BSk,
which are derived from
generalized Skyrme interaction functionals supplemented
with  several correction terms. Unlike APR and HHJ, the BSk
EOSs are unified. It means that they can be used not only for the
homogeneous nucleon-lepton matter in the stellar core, but
also in the crust.
The Skyrme parameters of the underlying energy-density
functionals were fitted
by \citet*{GorielyCP10} taking into account
experimental and theoretical constraints on nuclear
matter and neutron matter.
\citet{Potekhin-ea13} derived analytical
parametrizations of
three BSk EOSs
(BSk19, 20, and 21).

BSk21 reproduces the EOS labelled ``V18'' in \citet{LiSchulze}.
This EOS model is selected, because it provides
comfortably large $M_\mathrm{max}=2.28\,M_\odot$ to
accommodate observations,
is most consistent
with the experimental constraints (see Fig.~1 of
\citealt{Potekhin-ea13} and the discussion therein), and has a
powerful predictive ability for properties of heavy
neutron-rich nuclides \citep[e.g.,][]{Wolf-ea13}. A
comparison of the EOSs APR, HHJ(0.1,0.7), SC+HHJ, and BSk21
is given in Fig.~\ref{fig:EOS}.

We will use two neutron star models, with $M=1.4\,M_\odot$ and
$1.85\,M_\odot$ (Table \ref{tab:models}). The former is an example
of a star with the standard (not too strong) neutrino emission in
the core, mainly the modified Urca process in a non-superfluid star.
The latter model is an example of a star whose neutrino emission is
enhanced by the direct Urca (briefly Durca) process \citep{lpph91}
in a small inner kernel ($1.05 \times 10^{15}~\gcc <  \rho < 1.134
\times 10^{15}$ \gcc{} for SC+HHJ or $8.09 \times 10^{14} ~\gcc <
\rho < 9.98 \times 10^{14}$ \gcc{} for BSk21). For simplicity, we
consider non-superfluid neutron star models. Superfluidity would
affect the neutrino emission and heat capacity of the neutron
star core; it would further complicate our analysis. Detailed
studies of superfluid neutron stars with internal  heating can be
subject of a separate project.

\begin{table}
\caption[]{Four models (a)--(d) of heater positions $\rho_1 \leq
\rho \leq \rho_2$ in 1.4 and 1.85\,$M_\odot$ stars}
\label{tab:rho12}
\begin{center}
\begin{tabular}{cccc}
\hline 
  $M$:   & 1.4 \& 1.85\,$M_\odot$        &  $1.4~M_\odot$ &  $1.85~M_\odot$ \\
\hline
 label   & $\rho_1$ (\gcc) & $\rho_2$ (\gcc) &  $\rho_2$ (\gcc)  \\
\hline \hline
(a)  & $3.2 \times 10^{10}$ & $9.20 \times 10^{10}$ & $9.34 \times 10^{10}$ \\
(b)  & $3.2 \times 10^{11}$ & $1.60 \times 10^{12}$ & $1.60 \times 10^{12}$  \\
(c)  & $3.2 \times 10^{12}$ & $1.27 \times 10^{13}$ & $1.26 \times 10^{13}$ \\
(d)  & $3.2 \times 10^{13}$ & $5.47 \times 10^{13}$ & $5.39 \times 10^{13}$ \\
\hline
\end{tabular}
\end{center}

\end{table}

\begin{figure*}
\begin{minipage}{.75\linewidth}
\includegraphics[width=\linewidth]{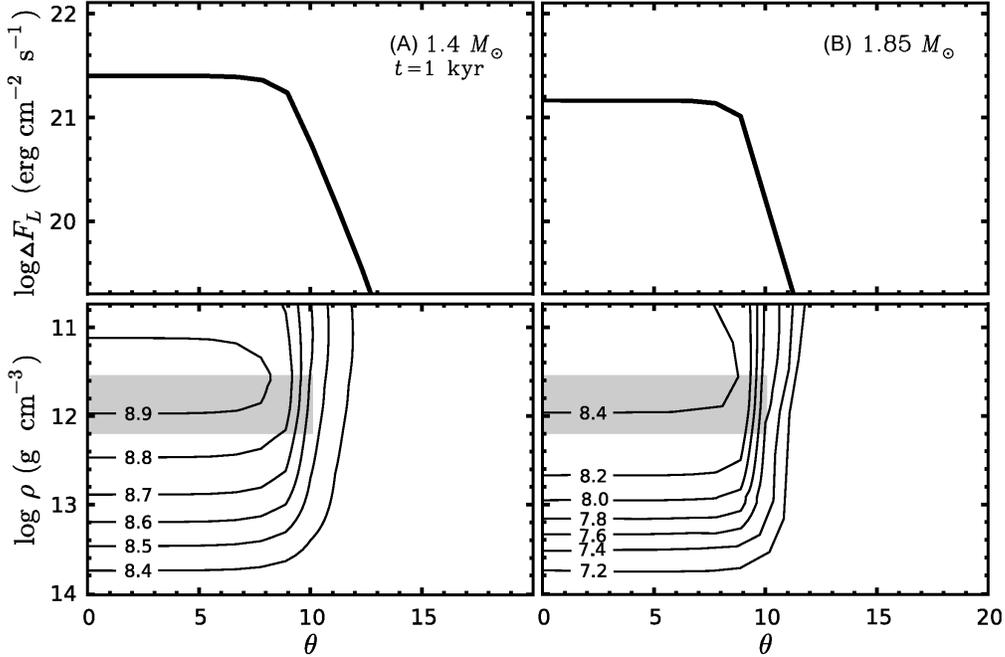}
\caption { {\it Upper panels}: Excess thermal flux density $\Delta F_L$
through the neutron star surface produced by the 2D heater
of type 2(b) in notations of Tables \ref{tab:H0} and \ref{tab:rho12},
with $\theta_0=10^\circ$, in the star of mass (A)
1.4\,$M_\odot$ (left) and (B) 1.85\,$M_\odot$ (right) at $t=1$~kyr,
as a function of polar angle $\theta$. {\it Lower panels}: Lines of
constant $\log_{10} T$ (numbers next to curves) within the star
on the plane $(\theta,\rho)$; shaded domains locate the heater.}
\label{fig:2D}
\end{minipage}
\end{figure*}

Following Papers~I and II, we introduce an internal phenomenological
heat source located in a layer at $\rho_1 \leq \rho \leq \rho_2$. The
heat rate $H$ [\rate] is taken in the form
\begin{equation}
     H=H_0\, \Theta(\rho)\, \exp(-t/\tau),
\label{H}
\end{equation}
where $H_0$ is the initial (age $t=0$) heat
intensity; $\Theta(\rho) \approx 1$ in the middle of density interval
$\rho_1<\rho<\rho_2$, and $\Theta(\rho)$ vanishes outside this
interval; and $\tau$ is the e-folding decay time of the heat
release. We treat
$H_0, \rho_1, \rho_2$, and $\tau$ as free input
parameters. In the 2D calculations, we also assume that $\Theta$ is
independent of $\theta$ at $\theta<\theta_0$ and
$\Theta(\rho,\theta) = 0$ at
$\theta \geq \theta_0$. Then the heater looks like a hot blob of angular
size $\theta_0$. The total heat power $W^\infty$ (erg~s$^{-1}$),
redshifted for a distant observer, is
\begin{equation}
   W^\infty(t)= \int \mathrm{d}V\, \mathrm{e}^{2\Phi}\, H,
\label{W}
\end{equation}
where $\mathrm{d}V$ is a proper volume element
(as before, redshifted quantities are marked by
the index $\infty$). In calculations, we used the smooth function $\Theta(\rho)$
as discussed in Paper~II.
An exact shape of $\Theta(\rho)$ in Eq.~(\ref{H})
is unimportant
for the surface temperature distribution,
provided the total heat power $W^\infty$
is fixed (Paper II).

For each $M$ we take four positions of
the heater (labelled ``(a)--(d)'')
with
$\rho_1$ and $\rho_2$
given in Table \ref{tab:rho12};
the bottom densities $\rho_2$
are chosen in such a way that all the heaters
(a)--(d) have the same power
$W^\infty$ at the same heat intensity
$H_0$ (in a star of given mass).
As seen from Table \ref{tab:rho12},
the values of $\rho_2$ for 1.4\,$M_\odot$ and 1.85\,$M_\odot$
stars are very close
(non-distinguishable in the figures presented below).

\begin{table}
\caption[]{Five levels 0,\ldots 4 of heat intensity $H_0$ used in
calculations}
\label{tab:H0}
\begin{center}
\begin{tabular}{ l  c c  c  c c}
\hline
Label & 0 & 1 & 2 & 3 & 4 \\
\hline
$H_0$ (erg s$^{-1}$ cm$^{-3}$) & ~0~ & $10^{18.5}$ &
$10^{19.5}$ & $10^{20.5}$ & $10^{21.5}$ \\
\hline
\end{tabular}
\end{center}

\end{table}

In numerical examples we will often use five levels of heat
intensity $H_0$ labelled as 0,\ldots,4. Level 0 corresponds
to no heating, whereas the other four levels refer to
progressively stronger heating (Table~\ref{tab:H0}). As in
Papers~I and II, we set $\tau=5 \times 10^4$ yr and consider
the case of $t \ll \tau$ (as discussed below, exact values
of $t$ and $\tau$ are unimportant for our analysis).
Therefore, well within the heater ($\rho_1 \ll \rho \ll
\rho_2$; $\theta < \theta_0$) at $t \ll \tau$ we have
$H(\rho,\theta,t) \approx H_0$.

 The total heat power $W^\infty$ is usually most important in
applications.  Nevertheless, it is $H_0$  which determines thermal
state of the heater, as has been analyzed in Papers I and II (e.g.
Fig. 5 in Paper I), and will be discussed here in Sects.\
\ref{s:regimes} and \ref{s:efficiency}.

\section{Calculations with the 2D code}
\label{s:2D}

We have applied the 2D code to simulations of cooling
neutron stars with the internal blob-like heater in the crust. We
have used the EOS SC+HHJ and varied neutron star mass (1.4\,$M_\odot$
and 1.85\,$M_\odot$; Table~\ref{tab:models}) as well as
the parameters of the heater ($H_0$, $\rho_1$, $\rho_2$,
$\theta_0$; Tables~\ref{tab:rho12}, \ref{tab:H0}).
We have calculated the density of the heat flux $F_L$
emergent from a local part of the surface.
It depends on $t$ because the star cools down.
However, it is instructive to introduce
the excess heat flux density
\begin{equation}
   \Delta F_L=F_L-F_{L0},
\label{e:dFL}
\end{equation}
where $F_{L0}$ is the heat flux for the star of the same age but
without any heater. The excess flux $\Delta F_L$ appears to be
robust, almost independent of neutron star age and cooling dynamics
(as long as $t \ll \tau$). It describes the quasi-stationary thermal
state of the star that is regulated by the heater itself and is
independent of the cooling history; hereafter, we present the
results obtained at $t=10^3$~yr. These results depend on the stellar
age as well as on the EOS model only slightly  (see, e.g., Fig.\
3 of Paper I, and Figs.\ 6 and 7 of Paper II, and discussion in
these papers).

An example of 2D calculations is given
in Fig.~\ref{fig:2D}. The
heater is placed in an upper part of the inner
crust: model (b) in Table \ref{tab:rho12}.
The heat intensity is
$H_0=10^{19.5}$~erg~cm$^{-3}$~s$^{-1}$
(level 2 in Table \ref{tab:H0}), and the heater's angular
size is $\theta_0=10^\circ$.
The left and right panels in Fig.~\ref{fig:2D} correspond to
the standard and fast coolers, respectively. The upper
panels show the excess thermal flux at the surface as a
function of the polar angle $\theta$. The bottom panels
present lines of constant temperature $T$ as a function of
$\rho$ and $\theta$. The hatched regions on the
$(\rho,T)$-plane are occupied by the heater. Here, the
temperatures and surface fluxes are not redshifted.

One can see that the internal temperature under the heater (near the
bottom of the inner crust) in the 1.85~$M_\odot$ star is drastically
lower than in the 1.4~$M_\odot$ star. This is because of much
stronger neutrino cooling (via the Durca process) from the core of
the more massive star. Nevertheless the geometries of the
temperature distributions, $T(\rho,\theta)$, and excess surface
emissions, $\Delta F_L(\theta)$, look similar for both
stars. One sees that the generated heat does not intend to spread
along the surface but propagates almost radially from the
heater up to the surface and down to the core. The excess heat flux
produces a hot spot as a direct projection of
the heater on the surface. The excess flux
$\Delta F_L(\theta)$ is almost constant from $\theta=0$ to
$\theta\approx\theta_0-2^\circ$ and exponentially decreases
at $\theta\gtrsim\theta_0$ with e-folding width
$\lesssim1^\circ$.

Similar results have been obtained by \citet{pr12}, who
modelled heat outflow in a magnetar outburst near a magnetic pole
with anisotropic heat conduction throughout the crust. The
anisotropy becomes important in a strong magnetic field,
which suppresses the electron heat transport across field
lines. Here we see that
the same radial heat propagation takes place even with isotropic
conduction in the crust without magnetic field.

The important outcome of the present 2D calculations is that
one can accurately model heat propagation  in a local part
of the heater using the 1D (radial) approximation. Note,
however, that this conclusion does not apply in a strong
magnetic field near the lines where the field is tangential
to the surface, as discussed in \citet{pcy07} and confirmed
in numerical calculations, e.g., by \citet*{PonsMG09}.
Nevertheless, this conclusion allows us to employ the local
approximation and use our standard 1D cooling code for
studying the main features of the heat transport from
blob-like heaters. We employ this approach in the rest of
this paper.

\section{Calculations with the 1D code}
\label{s:1D}

\subsection{Two regimes}
\label{s:regimes}

\begin{figure}
\centering
\includegraphics[width=.7\columnwidth]{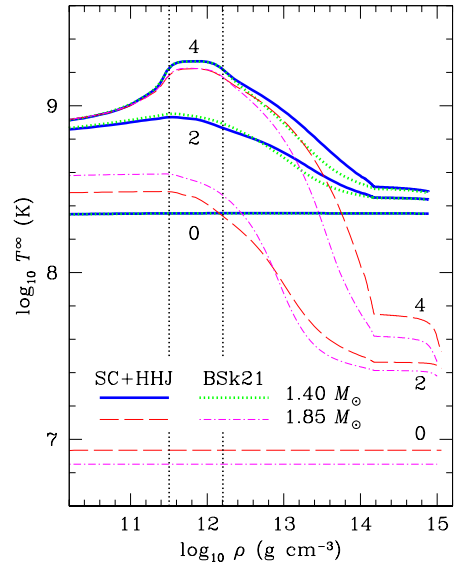}
\caption{
(Colour online)
Temperature profiles $T^\infty(\rho)$ in the $1.4\,M_\odot$ (thick solid lines)
and $1.85\,M_\odot$ (thin long-dashed lines) stars of age $t=1$ kyr
produced by a spherical heater (b) at two heat intensities (curves
2 and 4, Table \ref{tab:H0}) and without the heater (curves 0);
$\rho_1$ and $\rho_2$ (vertical dotted lines) are the same as in
Fig.~\ref{fig:2D}. In addition to the curves calculated for the SC+HHJ
model (solid and long-dashed lines), the curves for more
realistic BSk21 EOS (dotted and dot-dashed lines, respectively)
are plotted for comparison. See text for details.}
\label{fig:Ti}
\end{figure}

Fig.~\ref{fig:Ti} shows the redshifted temperature profiles
$T^\infty(\rho)=T(\rho) \,\mathrm{e}^\Phi$
inside the 1.4~$M_\odot$ and 1.85~$M_\odot$ stars
(``standard cooler'' and ``fast cooler'' in
Table~\ref{tab:models}) for the two EOS
models:  SC+HHJ and BSk21
(see Table~\ref{tab:models}).
It is $T^\infty(\rho)$ [not $T(\rho)$] that is constant
(independent of $\rho$) in isothermal regions of the star
with account for General Relativity.

For each star
and EOS
we show three $T^\infty(\rho)$ profiles
labelled 0,
2, and 4 according to Table~\ref{tab:H0}. The profiles 0 correspond to no heating
($H_0=0$). The profiles 2 and 4 are for the heater with
intensity $H_0=10^{19.5}$ and
$10^{21.5}$ erg~cm$^{-3}$~s$^{-1}$,
respectively, placed
at $3.2\times 10^{11} \leq \rho \leq 1.6\times
10^{12}$~g~cm$^{-3}$
(case (b) in Table~\ref{tab:rho12}).

The curves 0 correspond to ordinary cooling neutron stars and
decrease with the age $t$.
Without heating,
at $t=1$~kyr
the bulk of the star is already thermally relaxed
and $T^\infty$ stays nearly constant over the stellar
interior. In this case, the
internal temperature of the fast cooler is about 30 times lower
because of the
enhanced
neutrino cooling of the inner
kernel of the star.

Other curves (for the heated stars) are
different. The energy deposit in the crust destroys the thermally relaxed states and
makes $T^\infty(\rho)$ variable within the star.
The temperature profiles become mostly independent of $t$ (as long
as $t \ll \tau$), being supported by the heater. The stars stop to
cool down and reach (quasi)stationary states
(as plotted, e.g., in Figs.~2 and 3 of Paper~I). The hottest place in
the star is naturally the heater itself and its vicinity. The core
and the surface are colder, so that the generated heat is carried
away by thermal conduction to the surface and to the core. It is
also radiated away by neutrinos from different layers of the star.

As long as the heat intensity is not too strong ($H_0 \lesssim
10^{20}$ erg~cm$^{-3}$~s$^{-1}$, curves 2), the temperature
profiles in the fast cooler are noticeably lower than in
the standard cooler. This is again because of the stronger
neutrino emission in the core of the massive
star; strong neutrino
cooling in the core affects the heater
even if it is quite close to
the surface,
e.g.,
in the vicinity of the neutron drip
point.
If the heating is stronger ($H_0 \gtrsim 10^{20}$
erg~cm$^{-3}$~s$^{-1}$; curves 4 in Fig.~\ref{fig:Ti})
the situation is different. The temperature around the heater
ceases to depend on the neutrino emission in the
core, which manifests thermal decoupling of the heater and
the core, discussed below.

In each case, the results obtained for the two different
EOSs (SC+HHJ and BSk21) are qualitatively similar but
quantitatively different. The largest difference is observed
in the core of the fast cooler at the highest heating level,
where the
BSk21 model predicts considerably lower
temperature than the HHJ(0.1,0.7) model.
This is because the
Durca
threshold is lower for the
BSk21 EOS than for the
HHJ EOS (see
Table~\ref{tab:models}), which results in a more massive and
accordingly more efficient fast-cooling kernel of the
massive star.

\begin{figure}
\centering
\includegraphics[width=.99\columnwidth]{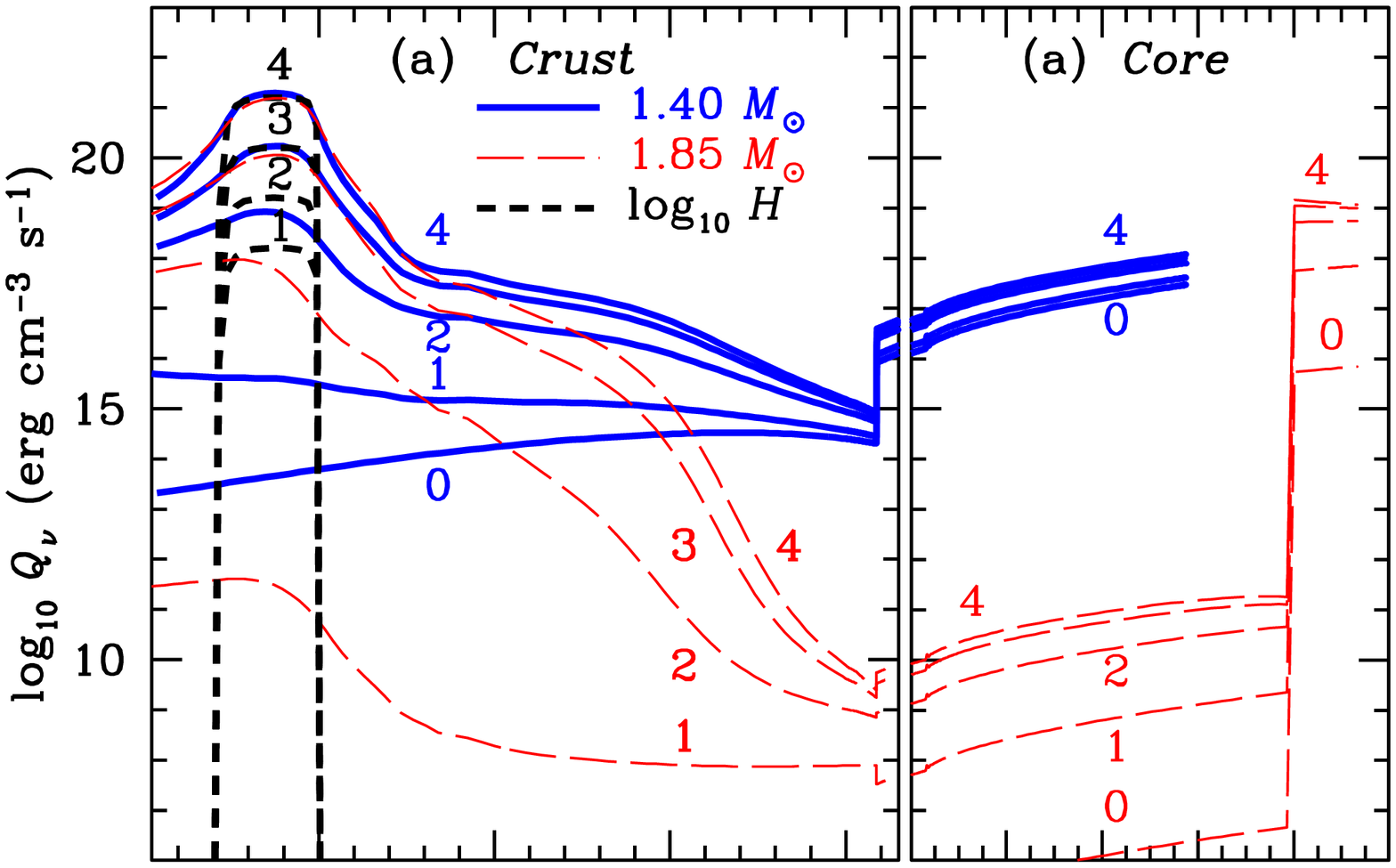}
\includegraphics[width=.99\columnwidth]{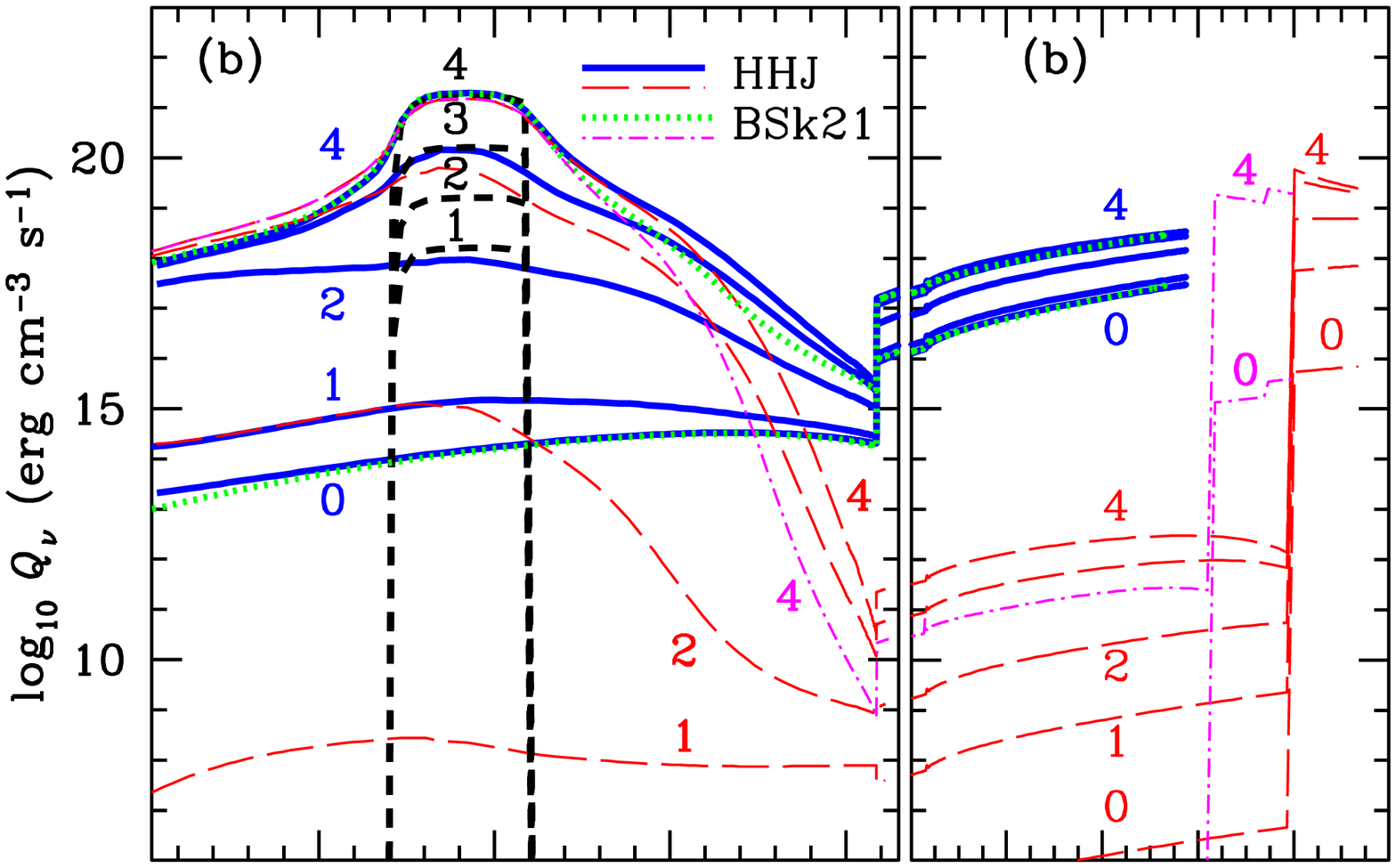}
\includegraphics[width=.99\columnwidth]{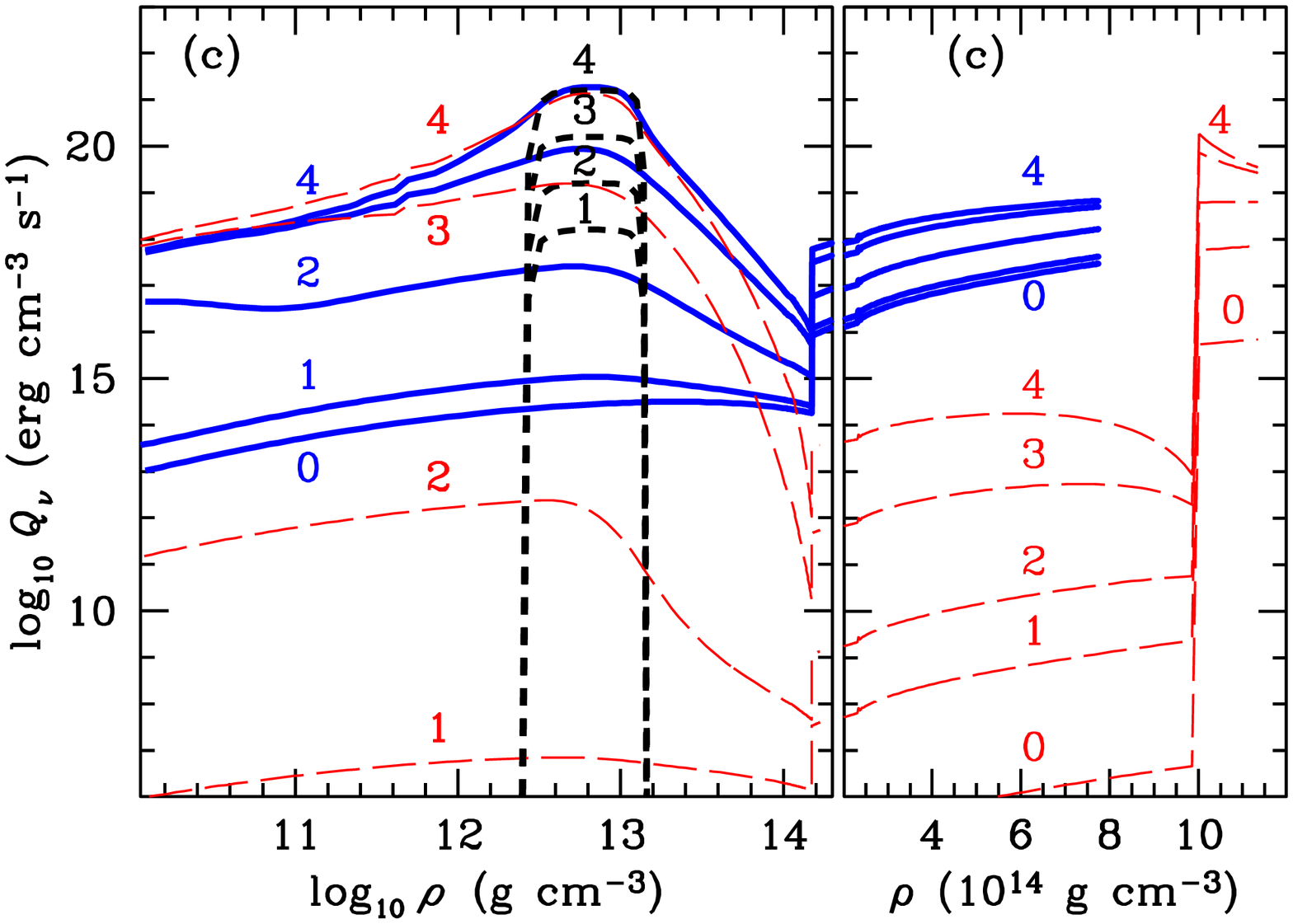}
\caption{ (Colour online) Logarithm of the neutrino emissivity
($Q_\nu$) versus density in the $1.4\,M_\odot$ star (thick solid
lines) and the $1.85 \, M_\odot$ star (thin long-dashed lines) 
with the SC+HHJ EOS at $t=1$ kyr for three positions of the heater
(a), (b) and (c) and five heat intensities 0,\ldots, 4 (Tables
\ref{tab:rho12} and \ref{tab:H0}). The left section of every panel
refers to the crust, with the density given in logarithmic scale;
the right section refers to the core, where $\rho$ is given in
linear scale (in units of $10^{14}$ g~cm$^{3}$). Short-dashed lines
give logarithm of the heat intensity, $\log H$. Additional dotted
and dot-dashed curves in the panels (b) show the results for
$M=1.4\,M_\odot$ and $1.85\,M_\odot$, respectively, calculated with
the BSk21 EOS for the heating levels 0 and 4.}
\label{fig:HQ}
\end{figure}

Fig.~\ref{fig:HQ} illustrates neutrino emission of the
heated neutron stars. It shows logarithm of the neutrino
emissivity $Q_\nu$ as a function of density throughout the
1.4\,$M_\odot$ and 1.85\,$M_\odot$ stars (thick solid and
thin long-dashed lines, respectively). The three panels (a),
(b) and (c) correspond to the three positions of the heater
in the crust (Table \ref{tab:rho12}). To visualize the
details,  the left section of each panel shows the crust
with the density in logarithmic scale, while the right
section shows the core
with the density in linear scale.
Visible jumps of the $Q_\nu$ curves at the crust-core
interface (near the right vertical axis of the left
panel) are due to the difference in neutrino emission
mechanisms in the crust and the core. The central density of
the 1.85\,$M_\odot$ star is higher than that of the
1.4\,$M_\odot$ star (Table \ref{tab:models}). Accordingly,
the $Q_\nu$ curves for the massive star are extended to
higher $\rho$. The large jumps of the $Q_\nu$ curves at $\rho\sim
10^{15}$ g~cm$^{-3}$ for the 1.85\,$M_\odot$ star are due to
the onset of the Durca neutrino emission in the central
kernel of the star.

Curves 0 correspond to ordinary cooling neutron stars. The
corresponding neutrino emission depends on
$t$.
Since the fast cooler is much
colder than the standard cooler (Fig.~\ref{fig:Ti}), its
neutrino emission in the outer core is much weaker at
densities before the Durca threshold, but becomes strongly
enhanced in the inner kernel after the threshold.

Curves 1--4 correspond to neutron stars heated from the crust. Their
neutrino emission is mainly ``frozen'' -- independent of $t$ (as
long as $t \ll \tau$), being primarily supported by the newly generated heat.

For each position of the heater and each $M$ we consider
five levels of heat intensity $H_0$ from zero to $10^{21.5}$
erg~cm$^{-3}$~s$^{-1}$ (curves 0,\ldots,4, Table \ref{tab:H0}). In each
case we plot also logarithm of the heat power
$H$. Fig.~\ref{fig:HQ} allows one to judge how much of the
input heat is transformed into neutrino emission and what
is the distribution of neutrino sources. Note that
in the preliminary publication \citep{kkpy12} similar curves
$Q_\nu(\rho)$ in Fig.~1(c) for $\log H_0$=18.5 and 19.5 were plotted
inaccurately.

In the middle panels (b) of Fig.~\ref{fig:HQ} we additionally plot
$Q_\nu(\rho)$ calculated for the BSk21 EOS model in the cases of no
heating and maximal heating. For the standard cooling
($M=1.4\,M_\odot$), the profiles $Q_\nu(\rho)$ for the two EOSs are
almost indistinguishable without heating and remain very close to
each other at the highest heating rate. For the fast cooling
($M=1.85\,M_\odot$), differences are more significant. First, the
jump at the Durca threshold is shifted to lower density according to
Table~\ref{tab:models} (see the right panel (b)). Second, the
neutrino emissivity becomes much lower at $\rho\gg10^{13}$~g
cm$^{-3}$, which is directly related to the lower $T^\infty$ in
Fig.~\ref{fig:Ti} due to the lower Durca threshold, as explained
above.

An analysis of Figs.~\ref{fig:Ti} and \ref{fig:HQ} (and other
figures presented below, as well as many other numerical results not
shown here) indicates the existence of two drastically different
regimes of energy outflow from the heater. These regimes are also
summarized in Table \ref{tab:regimes}.

(i) \emph{The conduction outflow regime} occurs at not too high heat
rates ($H_0 \lesssim 10^{20}$ erg~cm$^{-3}$~s$^{-1}$, curves 1 and 2
in Fig.~\ref{fig:HQ}) which produce not too high temperatures $T_\mathrm{h}
\lesssim 10^9$~K within the heater. In these cases,
$H\gg Q_\nu$ within the heater. The
heater's energy is mainly carried away by thermal
conduction. A
larger fraction of this energy sinks to the core and is emitted from
there by neutrinos. The thermal state of the crust in the heater and
its vicinity, as well as the thermal surface emission, are very
sensitive to the neutrino cooling in the core (modified or direct
Urca): there is \emph{a strong thermal coupling between the surface
and the core}.

(ii) \emph{The neutrino outflow regime} occurs at rather high heat
rates ($H_0 \gtrsim 10^{20}$ erg~cm$^{-3}$~s$^{-1}$, curves 3 and 4
in Fig.~\ref{fig:HQ}), i.e.{} at high temperatures $T_\mathrm{h} \gtrsim
10^9$~K within the heater. The heat power $H$ within the heater is
close to $Q_\nu$, which means that the heat is mainly radiated away by neutrinos
\emph{just in the heater}. A smaller fraction is carried away by
thermal conduction to the core, and only a tiny fraction of the heat
is conducted to the surface and radiated away as thermal emission.
The thermal states of the crust around the heater, as well
as thermal radiation from the surface, become insensitive to the neutrino
cooling in the core implying that \emph{the core is thermally decoupled from the
crust}.

\begin{table}
\caption[]{Two regimes of heat transport from the heater}
\begin{tabular}{lc@{\hspace{1ex}}cc@{\hspace{1ex}}c}
\hline
Regime  & $H_0$  & $T_\mathrm{h}$  &
Thermal & Coupling \\
  &  erg~cm$^{-3}$~s$^{-1}$ & K  & emission & to core \\
 \hline\hline
(i) Conduction  &    &     &  depend \\
~~outflow &  $\lesssim 10^{20}$  & $\lesssim 10^9$ &  on $H_0$ & yes \\
\hline
(ii) Neutrino   &         &       & \\
~~outflow  & $\gtrsim 10^{20}$ & $\gtrsim 10^9$ & saturated & no \\
\hline
\end{tabular}
\label{tab:regimes}
\end{table}

It is remarkable that the characteristic heat intensity
$H_0$ and heater's temperature $T_\mathrm{h}$, that separate regimes
(i) and (ii), are almost
independent of the heater's position in the crust.
Note, however, that
according to Paper~II,
$T_\mathrm{h}$  exceeds the characteristic value $\sim
10^9$~K  (at the same $H_0 \gtrsim 10^{20}$
erg~cm$^{-3}$~s$^{-1}$) if we shift the heater closer to the
surface (when $\rho_1 \lesssim 10^{10}$ g~cm$^{-3}$), but here
we do not analyze such shallow heaters. Note also
that when the heater is hot enough, the convective heat
transport may be initiated in the heater or its vicinity,
which we neglect for simplicity.

Because of different heat outflow regimes, one and the same neutron
star can show very different behavior if the heat power varies
within large limits. Let us remark also that our consideration of
thermal coupling/decoupling between the surface and the core
described above is strictly valid for spherical heaters. When the
heater looks like a blob and operates in the neutrino cooling
regime, the hot spot on the surface and the heater itself are
decoupled from the core. However, the surface layers outside the hot
spot can be thermally coupled to the core.

\begin{figure}
\includegraphics[width=\columnwidth]{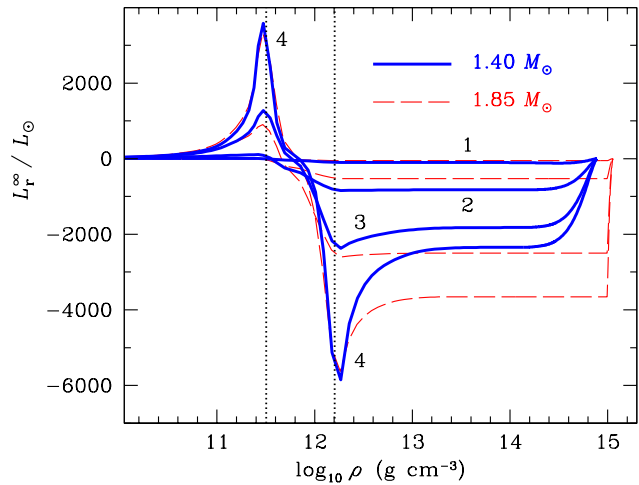}
\caption{ (Color online) Redshifted heat flux
$L^\infty(\rho)/L_\odot$ through spherical surfaces in the
1.4\,$M_\odot$ star (thick solid lines) and 1.85\,$M_\odot$ star
(thin dashed lines)  with the SC+HHJ EOS at $t=1$ kyr versus
$\rho$; the flux is directed outward (if positive) or inward (if
negative). The flux is produced by spherical heater (b) (whose
position is marked by dotted lines) at four heat intensities 1--4;
in all the cases the fluxes are relatively small at
$\rho=\rho_\mathrm{b}=10^{10}$~\gcc. }
\label{fig:L_r}
\end{figure}

Fig.~\ref{fig:L_r} shows the density dependence of
redshifted thermal heat-conduction flux through spherical
surfaces in the star (in units of solar luminosity
$L_\odot$) for the heater (b) (Table~\ref{tab:rho12}) with
four heat intensities $H_0$ (levels 1--4,
Table~\ref{tab:H0}). The thick solid lines are for the
1.4\,$M_\odot$ star, and the thin dashed lines are for the
1.85\,$M_\odot$ star.  The flux is positive when the heat
flows to the surface and negative when the heat is conducted
to the star's center. The curves are calculated at $t=1$
kyr, but they are virtually independent of $t$. The higher
$H_0$, the larger the heat flux. The largest fluxes occur at
the heater's boundaries $\rho_1$ and $\rho_2$; in these
places the fluxes are sensitive to the shape of the heat
power distribution $H(\rho)$.

Any flux vanishes at a certain  \emph{zero-flux surface} between the
heater's boundaries.  Above the zero-flux surface the heat  flows
outwards;  below this surface it flows to the core.  In the heat
outflow regime the zero-flux surface is closer to the outer
heater's boundary, $\rho_1$.  In the neutrino outflow regime it
shifts to the center of the heater, where the heat intensity is
maximal. For all heat intensities, maximum positive fluxes are
significantly lower than maximum negative ones. Accordingly, the
amount of the heat which flows to the star's surface is much smaller
than that which flows to the core.  In the outer core of the
1.85\,$M_\odot$ star the heat conduction flux is almost constant
because of low neutrino emission (Fig.~\ref{fig:HQ}); in the inner
kernel this flux rapidly decreases, because the Durca process
produces strong neutrino cooling there.

\subsection{How to warm up the surface}
\label{s:efficiency}

The main observational manifestation of the heater is the
thermal emission from the neutron star surface. Here we analyze the
ability of the heater to warm up the surface. The main obstacle for
warming up the surface is clear: the heat is mostly conducted
inside the star and radiated away by neutrinos.

\begin{figure}
\includegraphics[width=\columnwidth]{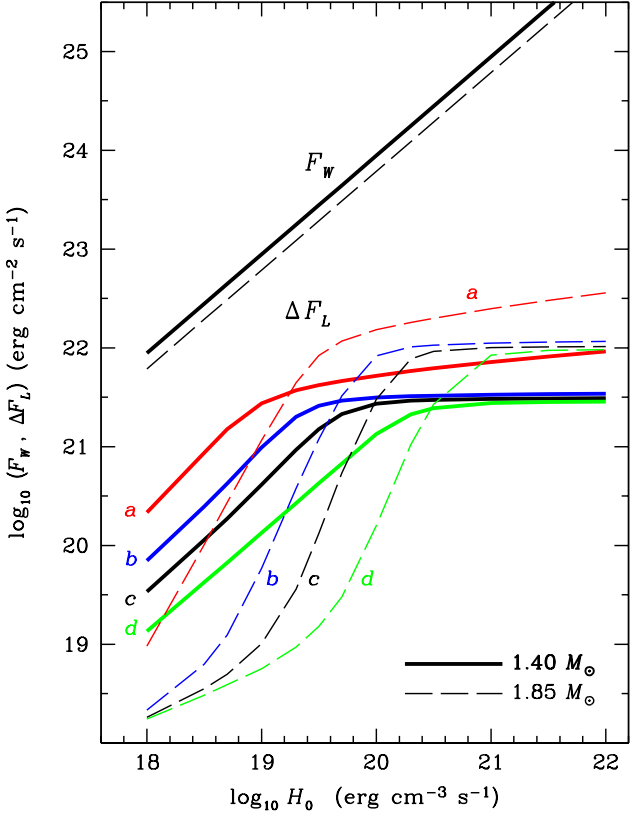}
\caption{ (Colour online) Generated heat flux $F_W$ and excess
surface flux $\Delta F_L$ for the standard cooler (solid lines) and
fast cooler (dashed lines) as functions of heat intensity $H_0$ for
different heater models. The upper straight lines of each type
(solid or dashed) show the total heat flux $F_{W}=W/(4\pi R^2)$ at
$t=1$ kyr (in a local reference frame) generated by the heater. The
other lines are excess heat fluxes $\Delta F_L$ through the stellar
surface from the heater of the same total power $W$ placed in four
regions (a)--(d).  The SC+HHJ EOS is used. See text for
details.}
\label{fig:dF_h}
\end{figure}

Fig.~\ref{fig:dF_h} shows the excess surface heat flux $\Delta F_L$
from the 1.4\,$M_\odot$ and 1.85\,$M_\odot$ stars  with  the
SC+HHJ EOS  (thick solid and thin dashed lines, respectively) 
and heaters of different intensities $H_0$ placed in four different
regions (a)--(d) of the crust (Table \ref{tab:rho12}). Recall
(Sect.~\ref{s:physics}) that we have chosen the widths of the
heaters (a)--(d) in such a way that they produce the same heat power
$W^\infty$ for the same heat intensity $H_0$. In Fig.~\ref{fig:dF_h}
we vary $H_0$ in a wide range. The upper line of each type (solid or
dashed) shows the flux $F_W=W/(4\pi R^2)$ (not redshifted for a
distant observer) as a function of $H_0$. It is determined by the
total non-redshifted heat power $W \approx W^\infty/(1-r_g/R)$,
where $r_g=2GM/c^2$ is the Schwarzschild radius of the star, and
$W^\infty$ is defined by Eq.~(\ref{W}). Other lines show the excess
thermal flux $\Delta F_L$ which reaches the surface from the heaters
(a)--(d). The $\Delta F_L/F_W$ ratio can be called the
\emph{efficiency of the heater} to warm up the surface  (with
$\Delta F_L/F_W=1$ if all the heater's energy could reach the
surface).

For example, the heater (b) with $H_0=3 \times 10^{19}$ \rate\
in the 1.4\,$M_\odot$ star would produce
$F_W \approx 2.8 \times 10^{23}$ and $\Delta F_L
\approx 2.6 \times 10^{21}$ erg~s$^{-1}$~cm$^{-2}$,
with $\Delta F_L/F_W \approx 10^{-2}$. The same heater in
the 1.85\,$M_\odot$ star
would generate
$F_W \approx 1.9 \times 10^{23}$ and $\Delta F_L
\approx 1.2 \times 10^{21}$ erg~s$^{-1}$~cm$^{-2}$,
with $\Delta F_L/F_W \approx 6 \times 10^{-3}$. Recall once more that
such quasi-stationary thermal states of these stars are determined by the heater.
Stars without heater would cool down. For instance, in the age interval from
$t=1$ to 10 kyr, the thermal flux $F_{L0}$ of the ordinary cooling 1.4\,$M_\odot$ star
would decrease from $2.1 \times 10^{20}$
to $8.4 \times 10^{19}$ erg~s$^{-1}$~cm$^{-2}$,
while for the 1.85\,$M_\odot$ star
it would decrease from $7.5 \times 10^{17}$
to $2.0 \times 10^{17}$ erg~s$^{-1}$~cm$^{-2}$.

\begin{figure}
\includegraphics[width=\columnwidth]{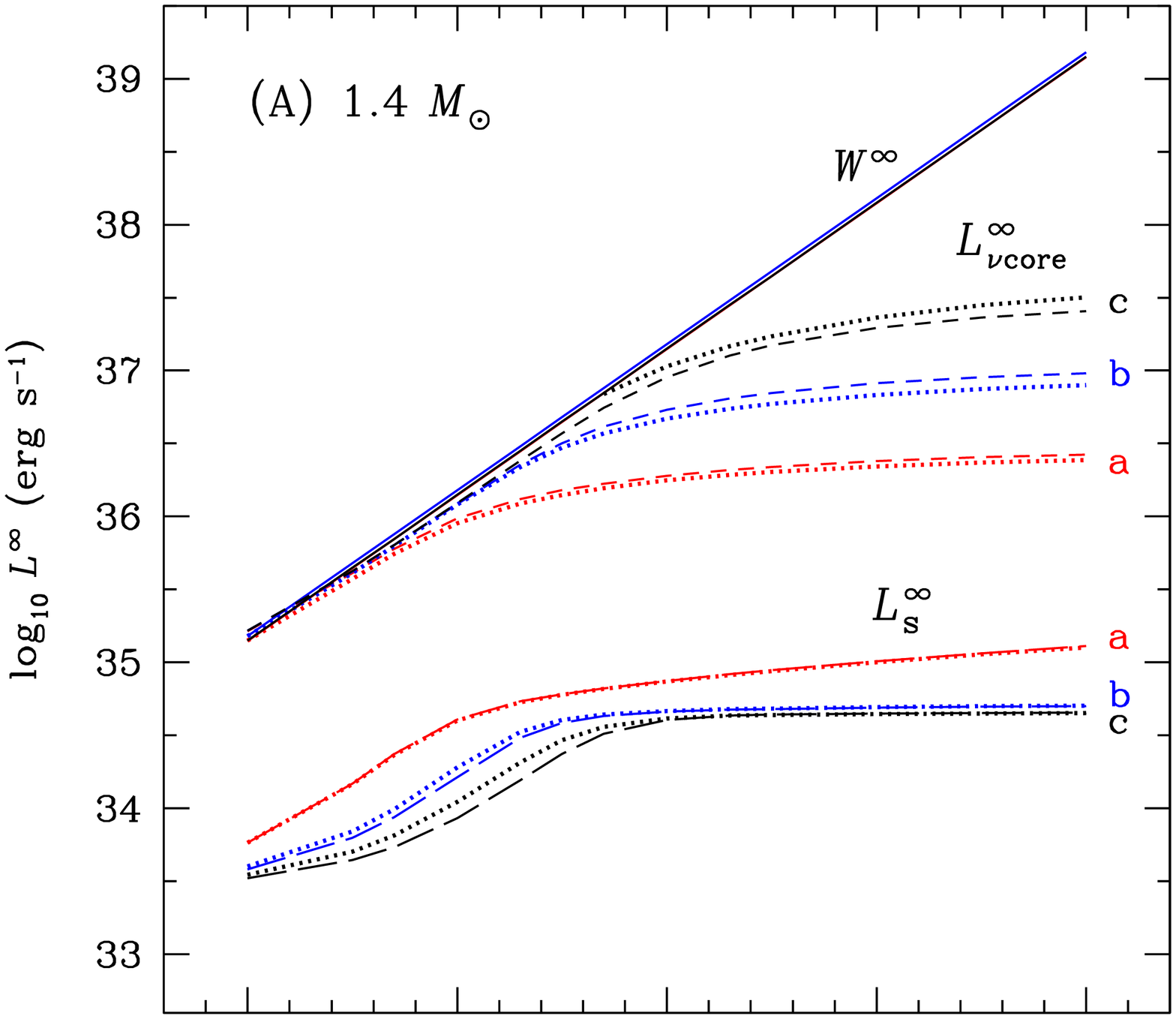}
\includegraphics[width=\columnwidth]{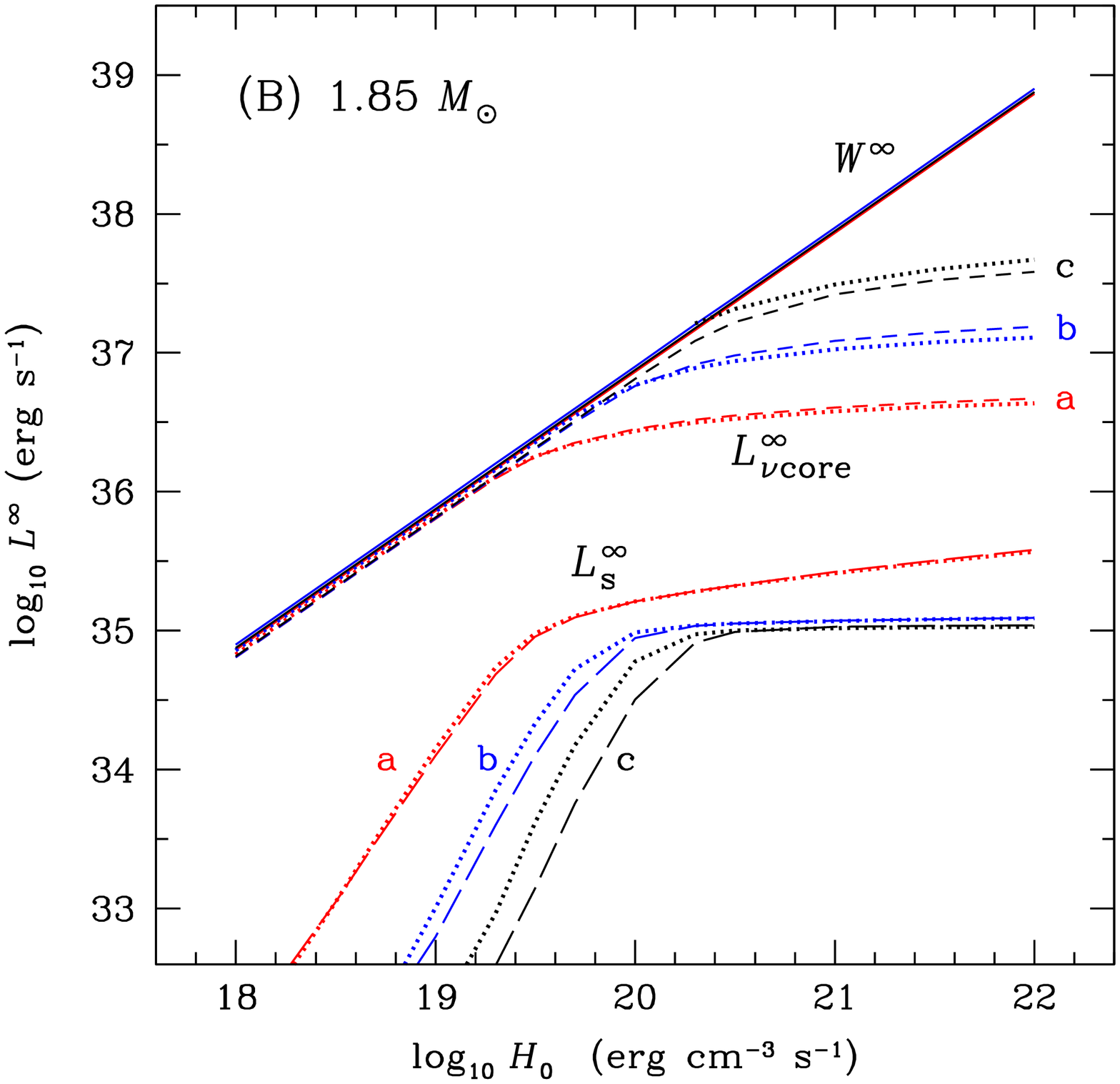}
\caption{
(Colour online)
Total redshifted heat power $W^\infty$
(thick solid lines),
the main part of
integrated redshifted neutrino luminosity of
the core $L_\mathrm{\nu core}^\infty$
(short-dashed lines),
provided by the crust-to-core heat flux (see text),
and surface
thermal luminosities $L_\mathrm{s}^\infty$ (long-dashed lines) versus
$\log H_0$ for the 1.4\,$M_\odot$ star [panel (A)] and the 1.85\,$M_\odot$
star [panel (B)] at the three heater's positions (a)--(c),
for the SC+HHJ EOS model.
The dotted lines that are close to the dashed ones show the
corresponding luminosities calculated with the
alternative model BSk21.}
\label{fig:WLL}
\end{figure}

Note that each group of curves in Fig.~\ref{fig:dF_h}, solid
or dashed, which represent the excess surface flux $\Delta
F_L$ as functions of $H_0$,  would look
exactly the same as functions of the total generated flux
$F_W$, because $F_W$ is a linear function of $H_0$ by
construction. In this way it is easy to see, how much of the
generated heat is radiated through the surface.

Fig.~\ref{fig:WLL} displays redshifted surface thermal
luminosities $L_s^\infty$ (long-dashed lines) emitted by the
1.4\,$M_\odot$ and 1.85\,$M_\odot$ stars [panels (A) and
(B), respectively] as functions of $H_0$, calculated for the
SC+HHJ EOS. The heaters are
placed in three regions (a)--(c).
In every case, a nearby dotted line
shows an analogous dependence for the BSk21 EOS. We plot
also the total redshifted heat power $W^\infty$ (solid
lines), and
the dominant part of
the integrated redshifted neutrino luminosity
from the star's core
$L^\infty_{\nu \rm core}$ (short-dashed lines) equal to
the total heat flux that passes from the crust to the core.
These curves
illustrate the contribution
in the heat balance
of the neutrino emission from the core.

Let us outline the main features of the conduction  outflow regime
(Table \ref{tab:regimes}).  In this regime the extra thermal
flux $\Delta F_L$ and the surface luminosity $L_s^\infty$ increase
with increasing $H_0$. This increase is sensitive to the physics of
the stellar core because of the thermal coupling between the core
and the surface. For the standard neutrino emission level in the
core (the modified Urca process) one approximately has $\Delta F_L
\propto H_0$ (Fig.~\ref{fig:dF_h}). One can show  that this
occurs for a warm heater, where the thermal conductivity weakly
depends on temperature (e.g., \citealt{gyp01}). For the enhanced
neutrino emission (the Durca process) at $H_0 \lesssim
10^{19}$~\rate\ the star is colder and $\Delta F_L$ is lower than
for the standard process. However, $\Delta F_L$ increases faster at
$H_0 \ga 10^{19}$~\rate\ and overcomes the standard values $\Delta
F_L$ at higher $H_0$.

 At lower $H_0$ in a massive cold star, the thermal conductivity
increases with lowering $T$ and destroys the relation $\Delta F_L
\propto H_0$. Note that we use the standard electron thermal
conductivity in the crust, neglecting possible effects of ion
impurities  \citep[e.g.,][]{pbhy99} and effects of distortion of electron wave functions due to interaction with the Coulomb lattice \citep{Chugunov12}. Both effects can strongly slow down the increase of the conductivity mention above and affect
thereby the relation between $\Delta F_L$ and $H_0$ at $H_0 \ll 10^{19}$
erg~cm$^{-3}$~s$^{-1}$ in massive stars.

In the conduction outflow regime $L_s^\infty$ increases with
$W^\infty$. The heater is divided into outer and inner parts, and
the heat is conducted to the surface from the outer part
(Fig.~\ref{fig:L_r}).

In the intermediate regime, at $H_0 \sim 10^{20}$
erg~cm$^{-3}$~s$^{-1}$, when the conduction outflow regime
transforms to the neutrino outflow regime, the flux $\Delta F_L$ becomes
nearly independent of the neutrino emission in the core,
but is mainly controlled
by neutrino emission in the crust.
This signifies the onset of thermal decoupling
between the surface and the core.

Finally we outline the neutrino outflow regime.  In this case the
flux $\Delta F_L$ and the luminosity $L_s^\infty$ saturate and
become almost independent of $H_0$  (and of the physics of the
core), except for the case (a) where the heat conduction to the
surface is more  competitive with the neutrino cooling. We expect
that moving  the heater even closer to the surface would increase
the surface thermal flux. However, for crustal heaters located in
the layers considered here, the generated surface thermal flux
weakly depends on the heat power and on the internal structure of
the star, while the efficiency of surface  photon emission $\Delta
F_L/F_W$ decreases with growing $H_0$. The heater can generate
enormous  amount of energy but it will be mostly radiated by
neutrinos and  will not increase the surface flux. This limiting
surface flux comes  from a thin outer layer of the heater. Making
the heater wider by extending it deeper within the crust will not
change the photon surface emission.

Therefore, there exist \emph{the maximum flux} $\Delta
F_L^\mathrm{max}$ that can emerge from a steady state heater in the
neutrino outflow regime. This maximum flux is almost the same   for
heaters (b)--(d) located deeper in the crust.   According to
Fig.~\ref{fig:dF_h}, $\Delta F_L^\mathrm{max}\approx 3 \times
10^{21}$ and $\Delta F_L^\mathrm{max}\approx 10^{22}$
erg~cm$^{-2}$~s$^{-1}$ for the 1.4\,$M_\odot$ and 1.85\,$M_\odot$
stars, respectively.   It is smaller than the Eddington flux
(the maximum steady radiation flux emergent through the neutron star
surface) by a factor of $\sim 10^{-4}-10^{-3}$.  The existence
of such a maximum radiation surface flux limited by neutrino
emission has been pointed out for magnetars (Papers I and II;
\citealt{pr12}), but it evidently exists for all neutron
stars.  In contrast to the papers mentioned above we calculate
it here assuming no anisotropy of heat transport in the neutron star
envelopes, particularly, in heat blanketing layers. For the heater
(a), which is closer to the surface, $\Delta F_L^\mathrm{max}$ is
higher.  We expect that moving the heater even closer to the surface
would further increase $\Delta F_L^\mathrm{max}$ making it closer to
the Eddington limit. These results imply that when a neutron
star radiates steadily at nearly Eddington luminosity, its radiation
cannot be powered by internal sources.

To summarize, the most efficient heater
would be intermediate between
the conduction and neutrino outflow regimes ($H_0 \sim 10^{20}$
erg~cm$^{-3}$~s$^{-1}$) and placed in the outer crust. It
would be uneconomical for the energy budget
to place the heater in the deep inner crust or to generate too much
heat ($H_0 \gg 10^{20}$ erg~cm$^{-3}$~s$^{-1}$). This conclusion,
already known for magnetars, remains valid for all
neutron stars.

In the neutrino outflow regime the efficiency of the heater in the
$1.85\,M_\odot$ star is somewhat higher than in the $1.4\,M_\odot$
star. This result seems counter-intuitive because the massive star
undergoes a very strong neutrino cooling. However, it is true,
because the more massive star has a  thinner crust, which
facilitates heat conduction to the surface.

The results of this section can be affected by strong
magnetic fields $B \gtrsim 10^{13}$~G and by chemical composition of
the heat blanketing envelope (as discussed, e.g., in Papers I and II).
The strongest effects are expected to occur
for most shallow heater's locations.

\section{Discussion}
\label{s:discuss}

In this section we outline the most important possible
manifestations of the internal heaters in neutron stars.

\subsection{Young cooling neutron stars}
\label{s:youngNSs}

Numerous simulations of young cooling neutron stars
(e.g., \citealt{lrpp94,yakovlevetal01}) demonstrate
the existence of quasi-stationary
thermal flux emergent from  neutron star
interiors. For
instance, Figs.~25 and 26 of \citet{yakovlevetal01} show cooling
curves of non-superfluid neutron stars of different masses with two
model EOSs in the core. They display the cooling ``as observed
from outside.'' There is a visible surface temperature drop
at $t\sim 10-10^2$ yr (depending on neutron star models and
microphysics input). It manifests the end of the initial thermal
relaxation inside cooling neutron stars. Snapshots of the
redshifted
temperature profiles $T^\infty(\rho)$ (of ``inside cooling'')
at different moments of time $t$
for two neutron stars of different masses are shown in Figs.~27
and 28 of \citet{yakovlevetal01}. Figures 25--28 of that
paper clearly demonstrate
the effects of temperature variations in a cooling neutron star
on its thermal photon emission.

Before the thermal relaxation ends, a star is strongly
non-isothermal inside. The crust is hotter than the core
because of lower neutrino emission in the crust. The
relaxation consists mainly in the core-crust equilibration.
It is accompanied by violent processes of non-uniform
neutrino cooling and heat conduction; the interior of the
star is highly non-isothermal, but the
surface temperature $T_\mathrm{s}$ in the period from
$\sim 0.1$ yr till the relaxation end stays wonderfully
constant, as if the star were thermally equilibrated, which
is definitely not the case!

Such quasistationary states of young cooling non-relaxed
neutron stars appear because
the temperature in some parts of the crust (Figs.~27 and 28
of \citealt{yakovlevetal01}) exceeds
$10^9$ K. This triggers the neutrino outflow regime and the
associated thermal decoupling. Then the quasistationary
thermal surface luminosity reaches the maximum
luminosity that the star can have (see Sect.~\ref{s:1D}).
Hot layers of the crust perform as powerful effective heaters. This
explains the results of numerous cooling simulations of
young neutron stars. Let us remark that the surface luminosity of
very young stars ($t \lesssim 0.1$ yr) is above the quasistationary
level and noticeably decreases with time.
This is because the very young neutron stars are far from
the steady state discussed in Sect.~\ref{s:1D}.

Note that some cooling simulations (e.g., \citealt{bgv04} and subsequent
publications based on similar microphysics) predict much stronger thermal emission
from surfaces of young neutron stars. These results are obtained
with non-standard physics of outer neutron-star layers.

\subsection{Neutron stars in soft X-ray transients}
\label{s:SXTs}

Internal heat sources operate also in transiently accreting
neutron stars in low-mass X-ray binaries (LMXBs; see, e.g.,
\citealt*{wdp13,tap13} and references therein). These objects
can be in active or quiescent states. In the active states,
neutron stars accrete matter from their low-mass companions
through accretion disks. The accretion strongly heats the
neutron star surface and triggers X-ray bursts in the
surface layers. Then  the neutron star is observed as a bright
X-ray source. Active states are followed by quiescent states
when the accretion is
quenched. Then X-ray luminosity decreases,
but the neutron star still shows noticeable thermal X-ray
emission indicating that it remains warm inside.

Quiescent thermal emission of transiently accreting neutron stars in
LMXBs is currently explained \citep{bbr98} by the deep crustal
heating of these stars \citep{hz90,hz08}. This heating operates over
the active states in the crustal matter compressed by newly accreted
material. The compression induces nuclear transformations
(absorption/emission of neutrons; electron captures; pycnonuclear
reactions) with release of $\sim1$--2 MeV per accreted nucleon, 
predominantly, in the inner crust.

Observations combined with theoretical models indicate
(e.g., \citealt{wdp13,tap13}) that the deep crustal
heating is insufficiently strong to endure the thermal
decoupling. All the sources remain in the conduction outflow
regime but behave in different ways.

First,
most of the sources
perform as quasi-stationary ones
(e.g., Aql X-1),  where the heater is not very strong or
operates for not too long, so that it does not violate
internal isothermality. The heater warms up the star during
the active states, and the heating is followed by the
cooling in the quiescent states. Such stars are thermally
inertial; heat gains and losses are thought to be balanced
over a few accretion cycles; the star reaches a
quasi-stationary state determined by crustal heating rate
(i.e., by the mass accretion rate) averaged over
$t\sim100$--1000 yr.

Second,
some sources (such as MXB 1659--298, KS 1731--260, EXO
0748--676, XTE J1701--462, IGR J17480--2446, MAXI J0556--332; see, e.g,
\citealt{degenaaretal13a,degenaaretal13b}, and references therein) can
be essentially nonstationary. In these cases the heater is strong or
operates for a sufficiently long time to overheat
the crust and violate the
thermal balance of the crust with the core.
After the accretion stops, the crust
starts to thermally equilibrate with the core, which is
manifested by
a surface temperature fall in the quiescent state over a few
months--years. It is actually the crust cooling observed in real
time. In contrast to the thermal relaxation in young neutron stars
(Sect.~\ref{s:youngNSs}), this relaxation
proceeds in the conduction outflow regime and
does not contain the stage of internal thermal decoupling.

\subsection{Magnetars and high-$B$ pulsars}
\label{s:magnetars}

Our results can help to interpret observations of soft gamma
repeaters (SGRs) and anomalous X-ray pulsars (AXPs), which are
thought to be magnetars, viz.{} neutron stars with superstrong
magnetic fields $B \gtrsim 10^{14}$~G (e.g.,
\citealt{m08,re11,m13,ok13}). The results can aslo be useful for
understanding the relations of the above sources to rotation powered
high-$B$ pulsars (e.g., \citealt{livingstone11,olausenetal13}). Let
us outline the physics of these objects, which is possibly affected
by internal heating.

SGRs and AXPs demonstrate slow rotation and large spindown rates
indicating they have very strong magnetic fields. There is
increasing evidence for the absence of any real difference between
AXPs and SGRs \citep*[e.g.,][]{GavriilKW02,m13}. SGRs/AXPs exhibit
large persistent thermal and non-thermal high-energy emission, X-ray
and gamma-ray bursts and flares (losing more energy than their
magnetic braking). This indicates wild processes of energy release
in their interiors and/or magnetospheres.

Moreover, AXPs/SGRs seem related to high-$B$ pulsars
\citep[e.g.,][]{Kaspi10,m13,Rea13}. The high-$B$ pulsars show
persistent thermal emission which is intermediate between magnetars
and standard radio pulsars, and, at least for one case (PSR
J1846--0258, \citealt{Gavriil08}), they demonstrate magnetar-like
outbursts. A high-$B$ pulsar can exhibit X-ray bursts and
then return to its initial state (e.g., \citealt{livingstone11}).

SGR/AXP-like activity is revealed even by some X-ray sources,
whose spin-down indicate lower fields $B\ll10^{14}$~G; this
may be a late manifestation of magnetar activity which is expected
to decay with age (e.g., \citealt{te13,Rea13,zhouetal14,reaetal14}
and references therein).

It seems that these features can be understood  assuming that
magnetized neutron stars possess persistent or variable internal heaters. When the
heaters are on, neutron stars can behave as SGRs/AXPs, but when the
heaters are off or weak, they behave as pulsars. Of course, this
internal activity can be closely related to the magnetospheric one (e.g.
\citealt{Beloborodov13}).

Because energy reservoirs for the heaters are limited, the heaters
should be economical (located not too deep in the crust and be not
too strong, Sect.~\ref{s:1D}). Such sources can produce thermal
decoupling between the neutron-star surface and the interior. Note
that the heater's efficiency can be higher in a more massive star
(with thinner crust), in a star with stronger magnetic field ($B
\gtrsim 10^{13}$~G) or in a star with heat blanketing envelope composed
of light elements (Papers I and II).

The heater's model may be like this.  If the heater is located in
the outer crust or near it, typical length-scales of pressure and
density variations are small and the electric conductivity is low
(especially if the heater is hot). Then the heater may be located in 
a special region, where non-linear MHD 
instabilities (triggered by crustal breaking or
magnetospheric activity) could take place. Here the Ohmic
decay
of electric currents can be strongly enhanced and produce the
required amount of heat (e.g.,
\citealt{kkpy12}, \citealt{viganoetal13}). 
The heaters may be
variable over months-years (appear, move, or almost disappear),
which can regulate long-term variability of magnetar activity. Our
results may help to develop a selfconsistent theory of
quasi-stationary states.
Another serious problem is to explain magnetar outbursts, their
origin and relaxation; there is a variety of ideas, 
e.g., \citealt{PernaPons11}, 
\citealt{PonsPerna11},   \citealt{ll12}, \citealt{viganoetal13}, 
and references therein.

An important problem is the energy delivery to the magnetic heater.
Magnetars lose too much energy, which  cannot be stored within one
heater's region. This energy can be accumulated in the internal
magnetic field of the star and then transported to the heater
(e.g., \citealt{viganoetal13}).
Evidently, the theory of
magnetar structure and evolution should be further elaborated.

\subsection{Neutron star mergers}
\label{s:mergers}

Merging neutron stars attract wide attention (e.g., \citealt{fr12}),
mainly because they are perspective objects to be detected by the new
generation of gravitational observatories (like advanced LIGO).
Gravitational signals from binary neutron star mergers are thought
to carry important information on the internal structure of neutron
stars.

Before neutron stars merge, they are likely heated by
hydrodynamical motions due to tidal interactions and
associated phenomena. One can treat this heating as produced
by internal heaters, so that the results of Sect.~\ref{s:1D}
can apply, at least qualitatively. The main outcome is that
after the internal temperature becomes sufficiently high in
certain layers of merging neutron stars, the neutrino
outflow regime starts to operate and govern the thermal
evolution of these layers. The thermal energy in these
layers will be efficiently carried away by neutrinos. A
disregard of neutrino emission in numerical simulations may
lead to inadequate physical  picture of merging neutron
stars.

\section{Conclusions}
\label{s:conclude}

We have studied the thermal surface radiation from neutron
stars with steady internal heaters. We have used our
new 2D code to consider blob-like heaters and our
standard 1D code to consider heaters located in
spherically symmetric layers. We have varied the sizes of
the heaters, as well as their power and position within the
crust. We have used neutron star models of two masses,
1.4\,$M_\odot$ and 1.85\,$M_\odot$. The 1.4\,$M_\odot$ star
has the standard neutrino emission from the core via the
modified Urca process, while the 1.85\,$M_\odot$ star has
the fast neutrino emission via the Durca process. We have
used two EOSs, SC+HHJ and BSk21. The first of them is based
on a simple energy-density function and serves for our
calculations in most cases. The second one is more
elaborated and more realistic;
it serves to
examine
the sensitivity of the results to
variations of EOSs.

Our main aim was to investigate how much energy of a heater can be
emitted through the surface as thermal radiation, and which
information on the heater and internal structure of neutron stars
can be inferred from observations of this radiation.

Our main conclusions are the following.

\begin{enumerate}

\item Comparison of 1D and 2D calculations reveals that
generated heat has no tendency to spread along the star's
surface. The heat mainly diffuses to the interior of the
star and is
carried away by neutrinos from there, but a small
fraction diffuses outwards and is emitted as thermal surface
radiation. The heater creates a hot spot, which is just the
projection of the heater onto the surface. Therefore, heat
propagation (excluding some special cases; Sect.~\ref{s:2D})
can be approximately studied with the local 1D
approximation.

\item The heater can operate in the two regimes. If its power is
not very strong, so that the temperature in the heater
$T_\mathrm{h} \lesssim 10^9$~K ($H_0 \lesssim 10^{20}$
erg~cm$^{-3}$~s$^{-1}$), then thermal transport within the
heater is mainly
conductive.
In this case the
surface emission can be
greatly reduced by the enhanced neutrino emission in the
stellar core of a massive star. On the other hand, it can intensified by
the growth of the heater's power.

\item If the heater is more powerful ($T_\mathrm{h} \gtrsim
10^9$~K; $H_0 \gtrsim 10^{20}$ erg~cm$^{-3}$~s$^{-1}$), its
energy is mainly carried away by neutrinos.  The surface
thermal radiation becomes independent  of the heater's power
and of the physics of the core; it is the maximum thermal
radiation which can be carried away from the heater of given
geometry by conduction and emitted through the stellar
surface.  In this regime, the surface becomes thermally
decoupled from the interior and even strong variations of
heater's power cannot significantly change the surface
emission.

\item The most economical heater, which transports to the
surface the maximum fraction of the released energy, should
be placed in the outer crust and be moderately strong ($H_0
\sim 10^{20}$ erg~cm$^{-3}$~s$^{-1}$) to avoid
non-economical neutrino cooling. Its efficiency can be still
higher in a more massive neutron star (with thinner crust),
in the presence of a superstrong magnetic field or in the
case where the blanketing envelope consists of light
elements.

\end{enumerate}

Some of these conclusions were previously drawn for
strongly magnetized neutron stars (e.g., Papers I and II).
Now we have shown that they are
pertinent to all neutron stars,
including non-magnetized ones. These conclusions are robust,
they do not depend on the concrete EOS we use.

We have outlined (Sect.~\ref{s:discuss}) possible applications of
the above results to young neutron stars, neutron stars in soft X-ray
transients, to magnetars and high-$B$ pulsars, as well as to merging neutron stars.
Other applications include, for instance, heating due to
viscous friction in the presence of differential rotation (e.g.,
\citealt{csy13}), slow chemical equilibration of the star in the
course of its evolution \citep{pr10}, thermal evolution of pulsars
after glitches.

It is important to account for the
neutrino outflow regime in hot neutron stars with strong heaters. Such a
heater drastically affects the heat transport mechanisms and
produces thermal decoupling of the heater from
deeper
regions of the star. We argue that this regime can be
realized in young cooling neutron stars
before the end of internal thermal relaxation,
in magnetars, and in merging neutron stars.
In this paper we have studied steady
state heaters. It would be interesting to extend the
analysis to variable heaters.

\section*{Acknowledgments}
 The authors are grateful to anonymous referee for careful
reading of the manuscript and useful comments. This work has been
partly supported by the RFBR (grant No.~14-02-00868-a) and by the
State Program ``Leading Scientific Schools of RF'' (grant NSh
294.2014.2).


\label{lastpage}

\end{document}